\documentclass[12pt]{article}

\usepackage{amssymb,amsmath,graphicx}

\newcommand{\alf}{\alpha}

\newcommand{\gam}{\gamma}

\newcommand{\zet}{\zeta}

\newcommand{\om}{\omega}

\newcommand{\abs}[1]{\left|#1\right|}
\newcommand{\ol}[1]{\overline{#1}}

\newcommand{\tq}{\widetilde{q}}

\begin{document}

\begin{center}
{\large A unified approach to autocorrelation of Frank, Chu, and Milewski sequences}
\end{center}

\begin{flushright}
Idris Mercer\\
Florida International University \\
\verb+imercer@fiu.edu+
\end{flushright}

\begin{abstract}
We construct a family of perfect polyphase sequences
that has the Frank sequences, Chu sequences, and
Milewski sequences as special cases.
This is not the most general construction of this type,
but it has a particularly simple form.
We also include some remarks about the acyclic autocorrelations
of our sequences.
\end{abstract}

\section{Introduction}

A {\bf complex sequence} of {\bf length} $N$ is an $N$-tuple
$$
\mathcal{X} = (\xi_0, \xi_1, \ldots, \xi_{N-1})
$$
where each $\xi_j$ is a complex number of modulus 1.
If furthermore each $\xi_j$ is a $D$th root of unity,
then the sequence is called a {\bf $D$-phase sequence},
or {\bf polyphase sequence} if we don't specify the value of $D$.
Throughout this article, lowercase Greek letters will denote complex numbers
and nonbold Latin letters will denote nonnegative integers, unless otherwise stated.

If $0 \le k \le N-1$, we define the {\bf acyclic autocorrelations} of $\mathcal{X}$ by
$$
\alf_k = \sum_{j=0}^{N-k-1} \ol{\xi_j} \xi_{j+k}
$$
and we define the {\bf cyclic autocorrelations} of $\mathcal{X}$ by
$$
\gam_k = \sum_{j=0}^{N-1} \ol{\xi_j} \xi_{j+k}
$$
where the bar denotes complex conjugation, and in the definition
of $\gam_k$, the addition in the subscript is mod $N$, or equivalently,
we assume $\xi_j$ is defined for $j \ge N$ by $\xi_{j+N}=\xi_j$.

Note that $\alf_k$ is a sum of $N-k$ terms, each of modulus 1.
Hence $\abs{\alf_k} \le N-k$.
Also note that if $1 \le k \le N-1$, we have $\gam_k = \alf_k + \ol{\alf_{N-k}}$.
It follows that if $\gam_k=0$, then
$\abs{\alf_k} = \abs{\alf_{N-k}} \le k$.

We can regard $\alf_k$ and $\gam_k$ as measuring resemblance between
the sequence $\mathcal{X}$ and a version of $\mathcal{X}$ that has been shifted by
$k$ positions (acyclically or cyclically respectively).
We have $\alf_0=\gam_0=N$, which we may call the {\bf trivial} autocorrelations.
Informally, we consider a sequence to be ``good'' if its nontrivial
autocorrelations are close to 0
(so it is ``uncorrelated'' with shifted versions of itself.)

How close to 0 can we make the cyclic autocorrelations,
and how close to 0 can we make the acyclic autocorrelations?
The first of those questions has an easier answer than the second.

We call $\mathcal{X}$ a {\bf perfect} sequence if $\gam_k=0$ for all $k\ne0$.
Many families of perfect sequences have been studied,
including Frank sequences \cite{FZ},
Chu sequences \cite{Chu},
and Milewski sequences \cite{Mil}.
There are perfect sequences of every length $N\ge2$.
For a good recent survey of perfect sequences,
see \cite[Section~4.1]{Sch16}.

We define the {\bf peak sidelobe level} (abbreviated ``PSL'') of $\mathcal{X}$ by
$$
\mathbf{P}(\mathcal{X}) = \max_{1\le k\le N-1} \abs{\alf_k},
$$
and we define the {\bf energy} of $\mathcal{X}$ by
$$
\mathbf{E}(\mathcal{X}) = \sum_{k=1}^{N-1} \abs{\alf_k}^2,
$$
which are two natural measures of the ``size'' of the acyclic autocorrelations.
A complex sequence with PSL at most 1
(i.e., with $\abs{\alf_k}\le1$ for all $k\ne0$)
is called a {\bf generalized Barker sequence}.
There exist generalized Barker sequences of all lengths $N\le70$
(see \cite{NC}), and it has been conjectured that they exist for all lengths.
See \cite[Question~4.10]{Sch16}
or \cite[p.~119]{Bor}.

A generalized Barker sequence of length $N$ has energy at most $N-1$.
But if we want an infinite family of complex sequences of increasing lengths $N$
that have small energy, the best known infinite families
(which are the Chu sequences and Frank sequences)
have energy growing like $O(N^{3/2})$. See \cite{Sch13}.

The {\bf merit factor} of a length $N$ complex sequence $\mathcal{X}$
is defined by
$$
\mathbf{F}(\mathcal{X}) = \frac{N^2}{2\mathbf{E}(\mathcal{X})}
$$
so asking for small energy is equivalent to asking for
large merit factor. The energy and merit factor are related to
the $L^4$ norm on the unit circle of the polynomial whose
coefficients are the $\xi_j$ (specifically, minimizing the energy
of the sequence is equivalent to minimizing the $L^4$ norm of the
polynomial). See \cite[Chapter~15]{Bor}.

We will construct a family of perfect polyphase sequences
that contains the Frank sequences, Chu sequences, and Milewski sequences
as special cases.
This is not the most general construction of this type.
In 1995, Mow provided a construction that
included all known infinite families of perfect polyphase sequences \cite{Mow}.
For more discussion of families of perfect polyphase sequences, see \cite{FD}.

Our family of sequences is not as general as Mow's,
but it includes several well-known families in one 
surprisingly simple form.
We will refer to our sequences as ``LM sequences'', where LM could stand
for ``like Mow'' but also names two integer parameters we use.

Some facts are already known about acyclic autocorrelation
of families of perfect sequences. Turyn \cite{Tur}
showed that the PSL of the Frank sequence of length $N$
is asymptotically equal to $(1/\pi)\sqrt{N}$.
Mow and Li \cite{ML} showed that the PSL of the Chu
sequence of length $N$ is asymptotically equal to $C\sqrt{N}$
for a slightly larger constant.
The current author \cite{Mer} showed that the energy
of the Chu sequence of length $N$ is bounded above by
$(8/3\pi^{3/2})N^{3/2}$; this was improved by Schmidt \cite{Sch13}
who showed that the energy of the Chu sequence of length $N$
is asymptotically equal to $(1/\pi)N^{3/2}$,
and the energy of the Frank sequence of length $N$
is asymptotically equal to $(2/\pi^2)N^{3/2}$.
It appears that less is known about the PSL or energy
of Milewski sequences. For a good summary of acyclic autocorrelation
of polyphase sequences, see \cite[Section~4.2]{Sch16}.

In this article, we do not say much about the acyclic autocorrelations
of our sequences. As consequences of the proofs that our sequences are
perfect sequences, it will follow that the Frank sequence of length $N$
has PSL at most $\sqrt{N/2}$, and the Chu sequence of length $N$ has
PSL at most $\sqrt{N}$, which are slightly weaker than known results.
The current author conjectures that there exist polyphase sequences of
all lengths $N$ whose energy grows like $o(N^{3/2})$.
Perhaps further study of acyclic autocorrelations of general families
of perfect sequences will prove this conjecture.

We note that if $\mathcal{X}$ is a $D$-phase sequence, then each $\xi_j$ can be written as
$\zet_D^{p(j)}$, where $\zet_D = e^{2\pi i/D}$ and $p(j)$ belongs to the integers mod $D$.
We can specify the sequence by specifying the
values of $p(j)$.

\section{LM sequences}

Throughout the rest of this article, $L$ and $M$ are positive integers
such that $L$ divides $M$, and $N$ denotes $LM$.
Since $N=1$ is uninteresting, we assume $M\ge2$.
We will construct a $2M$-phase sequence of length $N$,
informally consisting of $M$ ``blocks'' each of length $L$.
As before, $\zet_D$ means $e^{2\pi i/D}$.

For all nonnegative integers $j$, we define
$$
s_j = \left\lfloor \frac{j}{L} \right\rfloor \qquad \mbox{and} \qquad
t_j = j - L\left\lfloor \frac{j}{L} \right\rfloor.
$$
Then $j=s_jL+t_j$.
One can verify that $s_{j+N} = s_j + M$ and $t_{j+N} = t_j$.

{\bf Proposition 2.1.} Let $L,M$ be as above, and let $A$ be an
integer with the same parity as $LM$.
Define a function on the nonnegative integers as follows:
$$
p(j) = 2s_jt_j + Ls_j^2 + As_j.
$$
Then $p$ is an integer-valued function that satisfies $p(j+N) \equiv p(j)$ mod $2M$.

{\it Proof.} We have
\begin{align*}
p(j+N) & = 2s_{j+N}t_{j+N} + Ls_{j+N}^2 + As_{j+N} \\
& = 2(s_j+M)t_j + L(s_j+M)^2 + A(s_j+M) \\
& = 2s_jt_j + 2Mt_j + Ls_j^2 + 2LMs_j + LM^2 + As_j + AM \\
& = p(j) + 2M(t_j+Ls_j) + M(LM+A)
\end{align*}
and since $LM+A$ is even, this proves the proposition.

It follows that if we define, for all $j$,
$$
\xi_j = \zet_{2M}^{p(j)},
$$
then we have $\xi_{j+N} = \xi_j$ for all $j$.

{\bf Definition 2.2.} We define the {\bf LM sequence} of length $N=LM$ to be
$$
\mathcal{X} = (\xi_0,\xi_1,\ldots,\xi_{N-1})
$$
where $\xi_j=\zet_{2M}^{p(j)}$ and $p(j)$ is as defined previously.

The sequence depends on our choice of $L$, $M$, and $A$.
(Recall we must have $L|M$, and $A \equiv LM$ mod 2.)
We single out some important special cases.

{\bf Special case (i).} If $M$ is even, choose $A=M$,
and if $M$ is odd, choose $A=L$. We then have
$$
p(j) = \begin{cases}
2s_jt_j + Ls_j^2 + Ms_j & \mbox{if $M$ is even}, \\
2s_jt_j + Ls_j^2 + Ls_j & \mbox{if $M$ is odd}.
\end{cases}
$$
{\bf Special case (ii).} If $M$ is even, choose $A=0$,
and if $M$ is odd, choose $A=L$. We then have
$$
p(j) = \begin{cases}
2s_jt_j + Ls_j^2 & \mbox{if $M$ is even}, \\
2s_jt_j + Ls_j^2 + Ls_j & \mbox{if $M$ is odd}.
\end{cases}
$$
{\bf Special case (i)a.} Consider the subcase of special case (i) where
$M=L$. Then $N=M^2$, and we have
$$
p(j) = 2s_jt_j + Ms_j^2 + Ms_j = 2s_jt_j + Ms_j(s_j+1).
$$
Since $s_j(s_j+1)$ is even, we have $p(j) \equiv 2s_jt_j$ mod $2M$.
We then have
$$
\xi_j = \zet_{2M}^{p(j)} = \zet_{2M}^{2s_jt_j} = \zet_M^{s_jt_j}
$$
which means $\mathcal{X}=(\xi_0,\ldots,\xi_{N-1})$ is the sequence of
length $M^2$ defined by
$$
\xi_j = \xi_{s_jM+t_j} = \zet_M^{s_jt_j}
$$
which is equivalent to the Frank sequence of length $M^2$ as defined in \cite{FZ} or \cite[Section~4.1]{Sch16}.

{\bf Special case (ii)a.} Consider the subcase of special case (ii) where
$L=1$. Then $N=M$, and we have
$$
p(j) = \begin{cases}
2s_jt_j + s_j^2 & \mbox{if $M$ is even}, \\
2s_jt_j + s_j^2 + s_j & \mbox{if $M$ is odd}.
\end{cases}
$$
If $L=1$, then $s_j=\lfloor j/1 \rfloor=j$ and $t_j=j-1j=0$, so the above becomes
$$
p(j) = \begin{cases}
j^2 & \mbox{if $M$ is even}, \\
j^2 + j & \mbox{if $M$ is odd}.
\end{cases}
$$
This means $\mathcal{X}=(\xi_0,\ldots,\xi_{N-1})$ is the sequence of length $M$
defined by
$$
\xi_j = \zet_{2M}^{p(j)} = \begin{cases}
\zet_{2M}^{j^2} & \mbox{if $M$ is even}, \\[1ex]
\zet_{2M}^{j^2+j} = \zet_M^{(j^2+j)/2} & \mbox{if $M$ is odd},
\end{cases}
$$
which is equivalent to the Chu sequence of length $M$ as defined in \cite{Chu} or \cite[Section~4.1]{Sch16}.

{\bf Special case (ii)b.} Consider the subcase of special case (ii) where
$L=G^H$ and $M=G^{H+1}$, where $H>0$. Then $N=G^{2H+1}$, and we have
$$
p(j) = \begin{cases}
2s_jt_j + G^H s_j^2 & \mbox{if $G$ is even}, \\
2s_jt_j + G^H (s_j^2+s_j) & \mbox{if $G$ is odd}.
\end{cases}
$$
This means $\mathcal{X}=(\xi_0,\ldots,\xi_{N-1})$ is the sequence of length $G^{2H+1}$
defined by
$$
\xi_j = \xi_{s_jG^H+t_j} = \begin{cases}
\zet_{2M}^{2s_jt_j + G^H s_j^2} 
= \zet_{M}^{s_jt_j + \frac{G^H}{2} s_j^2} & \mbox{if $G$ is even}, \\[1ex]
\zet_{2M}^{2s_jt_j + G^H (s_j^2+s_j)}
= \zet_{M}^{s_jt_j + G^H\frac{s_j^2+s_j}{2}} &\mbox{if $G$ is odd},
\end{cases}
$$
or equivalently,
$$
\xi_j = \xi_{s_jG^H+t_j} =
\begin{cases}
\exp\big(\frac{2\pi i}{G^{H+1}}(s_jt_j+\frac{G^H}{2}s_j^2)\big)
& \mbox{if $G$ is even}, \\[1ex]
\exp\big(\frac{2\pi i}{G^{H+1}}(s_jt_j+G^H\frac{s_j^2+s_j}{2})\big)
& \mbox{if $G$ is odd},
\end{cases}
$$
which is equivalent to the Milewski sequence of length $G^{2H+1}$
as defined in \cite{Mil} or \cite[Section~4.1]{Sch16}.

\section{Useful lemmas}

As always, $\zet_D$ denotes $e^{2\pi i/D}$.
In this section, $x$ denotes a real number.

{\bf Lemma 3.1.} If $a$ and $k$ are any integers and $k$ is not a multiple of $D$, then
$$
\sum_{j=a}^{a+D-1} \zet_D^{kj} = 0.
$$
Proof: This sum of $D$ terms is invariant under multiplication by $\zet_D^k\ne1$.

Lemma. We have $\abs{1-e^{ix}}=2\abs{\sin\frac{x}{2}}$.

Proof: $\abs{1-e^{ix}}^2 = (1-\cos x)^2+\sin^2x = 2-2\cos x = 4\sin^2\frac{x}{2}$.

Lemma. If $0 < x \le\frac\pi2$ then
$\csc x \le \frac{\pi}{2x}$.

Proof: If $0 < x \le\frac\pi2$, we have
$\sin x \ge \frac{2}{\pi}x > 0$,
and taking reciprocals proves the lemma.

%Corollary. If $0<x\le\pi$ then
%$\csc\frac{x}2 \le \frac{\pi}{x}$.

Lemma. If $\om=e^{ix}\ne1$, and $a<b$ are integers, then
$$
\abs{\om^a+\om^{a+1}+\cdots+\om^{b-1}} \le \abs{\csc\frac{x}2}.
$$
Proof:
If $S = \om^a + \om^{a+1} + \cdots + \om^{b-1}$, we have $S-\om S = \om^a-\om^b$
and so
$$
\abs{S} = \frac{\abs{\om^a-\om^b}}{\abs{1-\om}}
= \frac{\abs{e^{iax}-e^{ibx}}}{\abs{1-e^{ix}}}
\le \frac{2}{\abs{1-e^{ix}}} = \frac{2}{2\abs{\sin\frac{x}{2}}}
= \abs{\csc\frac{x}{2}}.
$$
{\bf Corollary 3.2.} If $k$ is not a multiple of $D$, then taking
$\om=\zet_D^k=e^{i(2\pi k/D)}\ne1$, we get
$$
\abs{\sum_{j=a}^{b-1} \zet_D^{kj}} \le \abs{\csc\frac{\pi k}{D}}.
$$
If furthermore we have $0 < k < D$, this becomes
$$
\abs{\sum_{j=a}^{b-1} \zet_D^{kj}} \le \csc\frac{\pi k}{D}.
$$
Definition. For positive integers $k$ and $D$,
let $\delta(k,D)$ be the distance from $k$
to the nearest multiple of $D$.
So if $k$ belongs to an interval of the
form $[jD,(j+\frac12)D]$, then
$k = jD+\delta(k,D)$, and if $k$ belongs
to an interval of the form
$[(j-\frac12)D,jD]$, then
$k = jD-\delta(k,D)$.
For example, $\delta(19,12)=\delta(29,12)=5$.

Fact. The function
$f(x)=\abs{\csc\frac{\pi x}{D}}$
is periodic with period $D$ and is symmetric
on the interval $[0,D]$, i.e.\ $f(D-x)=f(x)$.

{\bf Corollary 3.3.}
If $k$ is not a multiple of $D$,
and $k' = \delta(k,D)$, then
$$
\abs{\csc\frac{\pi k}{D}} = \csc\frac{\pi k'}{D}
$$
and furthermore, since
$0<k'\le\frac{D}{2}$, we have
$$
\abs{\csc\frac{\pi k}{D}} \le \frac{D}{2k'}.
$$

\section{Autocorrelation of LM sequences}

Throughout this section, $\mathcal{X} = (\xi_0, \ldots, \xi_{N-1})$
is the LM sequence of length $N$ defined previously.
So $L$, $M$, $A$, and $p(j)$ are as defined previously, and the autocorrelations
$\gam_k$ and $\alf_k$ are sums of terms of the form
$$
\ol{\xi_j} \xi_{j+k} = \zet_{2M}^{p(j+k)-p(j)}.
$$
Suppose $1 \le k \le N-1$. We want to show $\gam_k=0$,
and we want bounds on the size of $\alf_k$.

Note that either $k$ and $N-k$ are both multiples of $L$,
or $k$ and $N-k$ are both nonmultiples of $L$.

{\bf Proposition 4.1.} Let $\mathcal{X}=(\xi_0,\ldots,\xi_{N-1})$
be the LM sequence of length $N=LM$ defined previously.
Suppose $1 \le k \le N-1$, and suppose $k$ and $N-k$ are both
multiples of $L$. Then the autocorrelations $\gam_k$ and $\alf_k$
satisfy $\gam_k=0$ and $\abs{\alf_k} \le \sqrt{N/2}$.

{\it Proof.}
If $k$ and $N-k$ are both multiples of $L$, let $k = qL$
and let $N-k = rL$. So $q+r=M$, and $1 \le q \le M-1$.
We break the sums $\gam_k$ and $\alf_k$ into sums of $L$ terms:
\begin{align*}
\gam_k & = \sum_{j=0}^{ML-1} \ol{\xi_j} \xi_{j+k} = \sum_{i=0}^{M-1} \sum_{j=iL}^{(i+1)L-1} \ol{\xi_j} \xi_{j+k}, \\
\alf_k & = \sum_{j=0}^{rL-1} \ol{\xi_j} \xi_{j+k} = \sum_{i=0}^{r-1} \sum_{j=iL}^{(i+1)L-1} \ol{\xi_j} \xi_{j+k}.
\end{align*}
CLAIM 1: If $iL \le j \le (i+1)L-1$ and $k=qL$ as above, then
$$
p(j+k)-p(j) = 2qj + Lq^2 + Aq
$$
which is of the form $2qj+c_1$ where $c_1$ is independent of $i$ and $j$.

Claim 1 is proved in the appendix.

We then have
\begin{align*}
\gam_k & = \sum_{i=0}^{M-1} \sum_{j=iL}^{(i+1)L-1} \zet_{2M}^{p(j+k)-p(j)}
= \zet_{2M}^{c_1} \sum_{i=0}^{M-1} \sum_{j=iL}^{(i+1)L-1} \zet_{2M}^{2qj} \\
& = \zet_{2M}^{c_1} \sum_{j=0}^{ML-1} \zet_{2M}^{2qj}
= \zet_{2M}^{c_1} \sum_{i=0}^{L-1} \sum_{j=iM}^{(i+1)M-1} \zet_{2M}^{2qj} \\
& = \zet_{2M}^{c_1} \sum_{i=0}^{L-1} \sum_{j=iM}^{(i+1)M-1} \zet_{M}^{qj}
= \zet_{2M}^{c_1} \sum_{i=0}^{L-1} 0 = 0
\end{align*}
where we have used Lemma 3.1 and the fact that $q$ is not a multiple of $M$.
We also have
$$
\alf_k = \sum_{i=0}^{r-1} \sum_{j=iL}^{(i+1)L-1} \zet_{2M}^{p(j+k)-p(j)}
= \zet_{2M}^{c_1} \sum_{i=0}^{r-1} \sum_{j=iL}^{(i+1)L-1} \zet_{2M}^{2qj}
= \zet_{2M}^{c_1} \sum_{j=0}^{rL-1} \zet_{2M}^{2qj}
$$
which implies
$$
\abs{\alf_k} = \abs{\sum_{j=0}^{rL-1}\zet_{2M}^{2qj}}
= \abs{\sum_{j=0}^{rL-1}\zet_{M}^{qj}} \le \csc\frac{\pi q}{M}
= \csc\frac{\pi(k/L)}{M} = \csc\frac{\pi k}{N}
$$
where we have used Corollary 3.2 and the fact that $0<q<M$.

Since $\gam_k=0$, we know $\abs{\alf_k}=\abs{\alf_{N-k}}$.
Therefore when bounding $\abs{\alf_k}$, it suffices to consider
$k \le N/2$.
If $1 \le k \le \sqrt{N/2}$, then $\abs{\alf_k} \le k \le \sqrt{N/2}$.
If $\sqrt{N/2} \le k \le N/2$, then
$$
\abs{\alf_k} \le \csc\frac{\pi k}{N} \le \frac{N}{2k} \le \frac{N}{2\sqrt{N/2}} = \sqrt{\frac{N}{2}}
$$
which completes the proof of Proposition 4.1.

{\bf Proposition 4.2.} Let $\mathcal{X}=(\xi_0,\ldots,\xi_{N-1})$
be the LM sequence of length $N=LM$ defined previously.
Suppose $1 \le k \le N-1$, and suppose $k$ and $N-k$ are both
nonmultiples of $L$. Then the autocorrelation $\gam_k$
satisfies $\gam_k=0$.
Furthermore, in the special case $L=M$,
the autocorrelation $\alf_k$
satisfies $\abs{\alf_k}\le M = \sqrt{N}$.

{\it Proof.}
If $k$ and $N-k$ are both nonmultiples of $L$, let $k = qL+k_1$
and let $N-k = rL+k_2$, where $1 \le k_1,k_2 \le L-1$ and $k_1+k_2=L$.
So $q+r=M-1$. % and $0 \le q \le M-1$.
We break the sums $\gam_k$ and $\alf_k$ into sums of $k_1$ terms
and sums of $k_2$ terms:
\begin{align*}
\gam_k & = \sum_{i=0}^{M-1} \sum_{j=iL}^{iL+k_2-1} \ol{\xi_j} \xi_{j+k}
+ \sum_{i=0}^{M-1} \sum_{j=iL+k_2}^{(i+1)L-1} \ol{\xi_j} \xi_{j+k}, \\
\alf_k & = \sum_{i=0}^{r} \sum_{j=iL}^{iL+k_2-1} \ol{\xi_j} \xi_{j+k}
+ \sum_{i=0}^{r-1} \sum_{j=iL+k_2}^{(i+1)L-1} \ol{\xi_j} \xi_{j+k}.
\end{align*}
CLAIM 2: If $iL \le j \le iL+k_2-1$ and $k=qL+k_1$ where $k_1,k_2$ are as above, then
$$
p(j+k)-p(j) = 2ik_1 + 2qj + 2qk_1 + Lq^2 + Aq
$$
which is of the form $2ik_1 + 2qj + c_2$ where $c_2$ is independent of $i$ and $j$.

CLAIM 3: If $iL+k_2 \le j \le (i+1)L-1$ and $k=qL+k_1$ where $k_1,k_2$ are as above, then
$$
p(j+k)-p(j) = -2ik_2 + 2\tq j - 2\tq k_2 + L\tq^2 + A\tq
$$
where $\tq=q+1$. This is of the form $-2ik_2+2\tq j+c_3$ where
$c_3$ is independent of $i$ and $j$.

Claims 2 and 3 are proved in the appendix.

We then have
\begin{align*}
\gam_k & = \sum_{i=0}^{M-1} \sum_{j=iL}^{iL+k_2-1} \zet_{2M}^{p(j+k)-p(j)}
+ \sum_{i=0}^{M-1} \sum_{j=iL+k_2}^{(i+1)L-1} \zet_{2M}^{p(j+k)-p(j)} \\
& = \zet_{2M}^{c_2} \sum_{i=0}^{M-1} \zet_{2M}^{2ik_1} \sum_{j=iL}^{iL+k_2-1} \zet_{2M}^{2qj}
+ \zet_{2M}^{c_3} \sum_{i=0}^{M-1} \zet_{2M}^{-2ik_2} \sum_{j=iL+k_2}^{(i+1)L-1} \zet_{2M}^{2\tq j} \\
& = \zet_{2M}^{c_2} \sum_{i=0}^{M-1} \zet_{2M}^{2ik_1} \sum_{j'=0}^{k_2-1} \zet_{2M}^{2q(iL+j')}
+ \zet_{2M}^{c_3} \sum_{i=0}^{M-1} \zet_{2M}^{-2ik_2} \sum_{j'=k_2}^{L-1} \zet_{2M}^{2\tq(iL+j')} \\
& = \zet_{2M}^{c_2} \sum_{i=0}^{M-1} \zet_{2M}^{2i(k_1+qL)} \sum_{j'=0}^{k_2-1} \zet_{2M}^{2qj'}
+ \zet_{2M}^{c_3} \sum_{i=0}^{M-1} \zet_{2M}^{2i(\tq L-k_2)} \sum_{j'=k_2}^{L-1} \zet_{2M}^{2\tq j'} \\
& = \zet_{2M}^{c_2} \cdot \sum_{i=0}^{M-1} \zet_{2M}^{2ik} \cdot \sum_{j'=0}^{k_2-1} \zet_{2M}^{2qj'}
+ \zet_{2M}^{c_3} \cdot \sum_{i=0}^{M-1} \zet_{2M}^{2ik} \cdot \sum_{j'=k_2}^{L-1} \zet_{2M}^{2\tq j'} \\
& = \zet_{2M}^{c_2} \cdot \sum_{i=0}^{M-1} \zet_{M}^{ik} \cdot \sum_{j'=0}^{k_2-1} \zet_{M}^{qj'}
+ \zet_{2M}^{c_3} \cdot \sum_{i=0}^{M-1} \zet_{M}^{ik} \cdot \sum_{j'=k_2}^{L-1} \zet_{M}^{\tq j'} \\
& = \zet_{2M}^{c_2} \cdot 0 \cdot \sum_{j'=0}^{k_2-1} \zet_{M}^{qj'}
+ \zet_{2M}^{c_3} \cdot 0 \cdot \sum_{j'=k_2}^{L-1} \zet_{M}^{\tq j'} = 0
\end{align*}
where we have used Lemma 3.1. (The hypotheses of Proposition 4.2 say that $k$
is not a multiple of $L$ and hence not a multiple of $M$.)
We also have, by similar manipulations,
\begin{align*}
\alf_k & = \sum_{i=0}^{r} \sum_{j=iL}^{iL+k_2-1} \zet_{2M}^{p(j+k)-p(j)}
+ \sum_{i=0}^{r-1} \sum_{j=iL+k_2}^{(i+1)L-1} \zet_{2M}^{p(j+k)-p(j)} \\
& = \cdots \\
& = \zet_{2M}^{c_2} \cdot \sum_{i=0}^{r} \zet_{M}^{ik} \cdot \sum_{j'=0}^{k_2-1} \zet_{M}^{qj'}
+ \zet_{2M}^{c_3} \cdot \sum_{i=0}^{r-1} \zet_{M}^{ik} \cdot \sum_{j'=k_2}^{L-1} \zet_{M}^{\tq j'}
\end{align*}
which implies
\begin{equation} \label{alfk}
\abs{\alf_k} = \abs{\sum_{i=0}^{r}\zet_{M}^{ik}} \abs{\sum_{j'=0}^{k_2-1}\zet_{M}^{qj'}} 
+ \abs{\sum_{i=0}^{r-1}\zet_{M}^{ik}} \abs{\sum_{j'=k_2}^{L-1} \zet_{M}^{\tq j'}}.
\end{equation}
Now suppose we are in the special case $L=M$.
Note that $0 \le q \le M-1$. If $q=0$, then $k < L$, so $\abs{\alf_k} < k < L = M$.
If $q=M-1$, then $r=0$, so $N-k < L$, so $\abs{\alf_k} < N-k < L = M$.
So assume $1 \le q \le M-2$, implying $2 \le \tq \le M-1$.
Then neither $q$ nor $\tq$ is a multiple of $M$.
Lemma 3.1 then gives us
\begin{align*}
\sum_{j'=0}^{L-1} \zet_M^{qj'} & = \sum_{j'=0}^{M-1} \zet_M^{qj'} = 0, \\
\sum_{j'=0}^{L-1} \zet_M^{\tq j'} & = \sum_{j'=0}^{M-1} \zet_M^{\tq j'} = 0,
\end{align*}
which then implies
\begin{align*}
\abs{\sum_{j'=0}^{k_2-1} \zet_M^{qj'}}
& = \abs{\sum_{j'=k_2}^{L-1} \zet_M^{qj'}}
\le \min\{k_1,k_2\}, \\
\abs{\sum_{j'=0}^{k_2-1} \zet_M^{\tq j'}}
& = \abs{\sum_{j'=k_2}^{L-1} \zet_M^{\tq j'}}
\le \min\{k_1,k_2\}.
\end{align*}
Next, observe that we have $\delta(k,M) = \delta(k,L) = \min\{k_1,k_2\}$.
If we let $k'=\min\{k_1,k_2\}$, then Corollaries 3.2 and 3.3 give us
\begin{align*}
\abs{\sum_{i=0}^r \zet_M^{ik}} & \le \abs{\csc\frac{\pi k}{M}} \le \frac{M}{2k'}, \\
\abs{\sum_{i=0}^{r-1} \zet_M^{ik}} & \le \abs{\csc\frac{\pi k}{M}} \le \frac{M}{2k'},
\end{align*}
Then inequality (\ref{alfk}) implies
$$
\abs{\alf_k} \le \frac{M}{2k'} \cdot k' + \frac{M}{2k'} \cdot k' = M,
$$
which completes the proof of Proposition 4.2.

In the case $L \ne M$, it is less clear how to bound $\abs{\sum\zet_M^{qj'}}$
and $\abs{\sum\zet_M^{\tq j'}}$. We can bound them by $L$, but that does not
make it obvious whether $\abs{\alf_k}$ can be bounded by a multiple of $\sqrt{N}$.
A more careful analysis may be needed.

In summary, Propositions 4.1 and 4.2 together imply that every LM sequence
satisfies $\gam_k=0$ for all $k\ne0$, i.e.\ every LM sequence is a perfect sequence.
In the special case $L=1$ (which includes the Chu sequences),
$k$ is always a multiple of $L$, so Proposition 4.1
always applies, and the LM sequence has PSL at most $\sqrt{N/2}$.
In the special case $L=M$ (which includes the Frank sequences),
the LM sequence has PSL at most $\sqrt{N}$.

\section{Appendix}

As before, $L$ and $M$ are positive integers satisfying $L|M$,
and $A \equiv LM$ mod 2.
For all nonnegative integers $j$,
we define
\begin{align*}
s_j & = \left\lfloor \frac{j}{L} \right\rfloor, \\
t_j & = j - L \left\lfloor \frac{j}{L} \right\rfloor, \\
p(j) & = 2s_jt_j + Ls_j^2 + As_j.
\end{align*}
Proof of Claim 1.
If $iL \le j \le (i+1)L-1$ and $k=qL$, then
$$
(i+q)L \le j+k \le (i+q+1)L-1,
$$
so we have
\begin{align*}
s_j & = \left\lfloor \frac{j}{L} \right\rfloor = i, \\
s_{j+k} & = \left\lfloor \frac{j+k}{L} \right\rfloor = i+q, \\
t_j & = j - Li, \\
t_{j+k} & = (j+k) - L(i+q) = j - Li = t_j, \\
s_{j+k} - s_j & = q, \\
s_{j+k}^2 - s_j^2 & = (i+q)^2 - i^2 = 2iq + q^2, \\
s_{j+k}t_{j+k} - s_jt_j & = (s_{j+k}-s_j)t_j = q(j-Li).
\end{align*}
So then
\begin{align*}
p(j+k)-p(j) & = 2(s_{j+k}t_{j+k}-s_jt_j) + L(s_{j+k}^2-s_j^2) + A(s_{j+k}-s_j) \\
& = 2q(j-Li) + L(2iq+q^2) + Aq \\
& = 2qj + Lq^2 + Aq.
\end{align*}

Proof of Claim 2.
If $iL \le j \le iL+k_2-1$ and $k=qL+k_1$, and $k_1,k_2$ are positive integers
satisfying $k_1+k_2=L$, then
$$
(i+q)L + k_1 \le j+k \le (i+q+1)L - 1,
$$
so we have
\begin{align*}
s_j & = \left\lfloor \frac{j}{L} \right\rfloor = i, \\
s_{j+k} & = \left\lfloor \frac{j+k}{L} \right\rfloor = i+q, \\
t_j & = j - Li, \\
t_{j+k} & = (j+k) - L(i+q) = j - Li + k_1, \\
s_{j+k} - s_j & = q, \\
s_{j+k}^2 - s_j^2 & = (i+q)^2 - i^2 = 2iq + q^2, \\
s_{j+k}t_{j+k} - s_jt_j & = (i+q)(j-Li+k_1)-i(j-Li) \\
& = ik_1 + qj - qLi + qk_1.
\end{align*}
So then
\begin{align*}
p(j+k)-p(j) & = 2(s_{j+k}t_{j+k}-s_jt_j) + L(s_{j+k}^2-s_j^2) + A(s_{j+k}-s_j) \\
& = 2(ik_1+qj-qLi+qk_1) + L(2iq+q^2) + Aq \\
& = 2ik_1 + 2qj + 2qk_1 + Lq^2 + Aq.
\end{align*}

Proof of Claim 3.
If $iL+k_2 \le j \le (i+1)L-1$ and $k=qL+k_1$,
and $k_1,k_2$ are positive integers satisfying $k_1+k_2=L$, then
$$
(i+q+1)L \le j+k \le (i+q+1)L + k_1 - 1.
$$
Let $\tq = q+1$.
Then we have
\begin{align*}
s_j & = \left\lfloor \frac{j}{L} \right\rfloor = i, \\
s_{j+k} & = \left\lfloor \frac{j+k}{L} \right\rfloor = i+\tq, \\
t_j & = j - Li, \\
t_{j+k} & = (j+k) - L(i+q+1) = j - Li - k_2, \\
s_{j+k} - s_j & = \tq, \\
s_{j+k}^2 - s_j^2 & = (i+\tq)^2 - i^2 = 2i\tq + \tq^2, \\
s_{j+k}t_{j+k} - s_jt_j & = (i+\tq)(j-Li-k_2) - i(j-Li) \\
& = -ik_2 + \tq j - \tq Li - \tq k_2.
\end{align*}
So then
\begin{align*}
p(j+k)-p(j) & = 2(s_{j+k}t_{j+k}-s_jt_j) + L(s_{j+k}^2-s_j^2) + A(s_{j+k}-s_j) \\
& = 2(-ik_2 + \tq j - \tq Li - \tq k_2) + L(2i\tq + \tq^2) + A\tq \\
& = -2ik_2 + 2\tq j - 2\tq k_2 + L\tq^2 + A\tq.
\end{align*}

\bibliographystyle{amsplain}

\end{document}